\documentclass[%
reprint,
superscriptaddress,
nofootinbib,
amsmath,amssymb,
aps,
prx,
]{revtex4-2}

\usepackage{graphicx}
\usepackage{dcolumn}
\usepackage{bm}
\usepackage{lipsum}
\usepackage{float}
\usepackage{xcolor}
\usepackage{soul}
\usepackage{hyperref}
    \hypersetup{
        pdftitle={Interaction-driven quantum phase transitions between topological and crystalline orders of electrons},
        colorlinks=true,
        linkcolor=blue,
        anchorcolor=blue,
        citecolor=blue,
        urlcolor=blue,
    }

\newcommand{\vect}[1]{{\bf #1}}
\begin{document}

\title{Interaction-driven quantum phase transitions between topological and crystalline orders of electrons}

\author{André Haug}
\altaffiliation{These authors contributed equally to this work.}
\affiliation{Department of Condensed Matter Physics, Weizmann Institute of Science, Rehovot, 76100, Israel.}
\author{Ravi Kumar}
\altaffiliation{These authors contributed equally to this work.}
\affiliation{Department of Condensed Matter Physics, Weizmann Institute of Science, Rehovot, 76100, Israel.}
\author{Tomer Firon}
\affiliation{Department of Condensed Matter Physics, Weizmann Institute of Science, Rehovot, 76100, Israel.}
\author{Misha Yutushui}
\affiliation{Department of Condensed Matter Physics, Weizmann Institute of Science, Rehovot, 76100, Israel.}
\author{Kenji Watanabe}
\affiliation{Research Center for Functional Materials, National Institute for Materials Science, 1-1 Namiki, Tsukuba 305-0044, Japan.}
\author{Takashi Taniguchi}
\affiliation{International Center for Materials Nanoarchitectonics, National Institute for Materials Science, 1-1 Namiki, Tsukuba 305-0044, Japan.}
\author{David F. Mross}
\affiliation{Department of Condensed Matter Physics, Weizmann Institute of Science, Rehovot, 76100, Israel.}
\author{Yuval Ronen}
\email{yuval.ronen@weizmann.ac.il}
\affiliation{Department of Condensed Matter Physics, Weizmann Institute of Science, Rehovot, 76100, Israel.}

\date{\today}

\begin{abstract}
Topological and crystalline orders of electrons both benefit from enhanced Coulomb interactions in partially filled Landau levels. In bilayer graphene (BLG), the competition between fractional quantum Hall liquids and electronic crystals can be tuned electrostatically. Applying a displacement field leads to Landau-level crossings, where the interaction potential is strongly modified due to changes in the orbital wave functions. Here, we leverage this control to investigate phase transitions between topological and crystalline orders at constant filling factors in the lowest Landau level of BLG. Using transport measurements in high-quality hBN-encapsulated devices, we study transitions as a function of displacement field near crossings of $N=0$ and $N=1$ orbitals. The enhanced Landau-level mixing near the crossing stabilizes electronic crystals at all fractional fillings, including a resistive state at $\nu = \frac{1}{3}$ and a reentrant integer quantum Hall state at $\nu = \frac{7}{3}$. On the $N=0$ side, the activation energies of the crystal and fractional quantum Hall liquid vanish smoothly and symmetrically at the transition, while the $N=1$ transitions out of the crystal appear discontinuous. Additionally, we observe quantized plateaus forming near the crystal transition at half filling of the $N=0$ levels, suggesting a paired composite fermion state stabilized by Landau level mixing.
\end{abstract}

\maketitle

\section*{Introduction}
A two-dimensional electron liquid in a magnetic field exhibits the classical Hall effect, where the Hall conductance $\sigma_{xy}$ depends only on the electron density and the magnetic field, independently of interactions. However, breaking Galilean invariance can decouple $\sigma_{xy}$ from these parameters. In the well-known quantum Hall effect \cite{Haldane_fqh_1983,Halperin_fqh_1983,Laughlin_fqh_1983,Jain_composite_2007}, weak disorder stabilizes plateaus of constant $\sigma_{xy}$ around specific ratios of density and magnetic field. A more dramatic manifestation of Galilean symmetry breaking arises when Coulomb interactions drive electrons into a Wigner crystal (WC) \cite{Wigner.1934}. The spontaneous breaking of translational symmetry in this phase enables the Hall conductance to deviate significantly from its classical value, even in the absence of disorder \cite{Kim.2021.Quantum}.

WCs with classical symmetry-breaking order compete against fractional quantum Hall (FQH) states exhibiting topological order at non-integer values of the filling factor $\nu$~\cite{Yoshioka_Charge_1979, Yoshioka_Ground_1983, Fogler_Laughlin_wigner_1997, Goerbig_Competition_2004, Chen.2019.Competing, Dora.2023}. Both types of phases benefit from enhanced Coulomb interactions as the kinetic energy is quenched in a magnetic field. In the lowest Landau level, WCs arise at very low fillings or, in the form of reentrant states, close to unit filling. They have been reported in multiple quantum Hall platforms, including GaAs, AlAs, ZnO, and graphene systems \cite{Andrei.1988.Observation, Goldman.1990, Jiang.1990, Santos.1992.Observation, Santos.1992.Effect, Kozuka.2011, maryenko2018composite, Zhou.2020, Rosales.2021.Competition, Singh.2024.Developing, Tsui.2024, Seiler.2024.Signatures}. Tuning the filling factor results in phase transitions from WCs to FQH states, which are governed by localization due to disorder. The bulk consists of WC and FQH puddles, leading to a network of edge channels that percolate at the transition point.

A fundamentally different transition between FQH and WC, driven by interactions instead of disorder, can occur at constant filling \cite{Apalkov.2010, Jolicoeur_Competing_2023}. Precisely at specific, rational values of $\nu$ such as $\frac{1}{3}$ or $\frac{2}{5}$, FQH liquids and WCs can both be stabilized even without disorder. The balance between these phases then depends on the interaction potential, which is affected by virtual excitations of electrons into nearby Landau levels. Several theoretical studies predict that Landau-level mixing (LLM) promotes WCs \cite{Zhao_Crystallization_2018, Jolicoeur_Competing_2023, Teng_Solving_2024}. A recent experimental study in AlAs quantum wells observed WCs in low-density samples and FQH states in higher-density samples at the same filling factor \cite{Csathy.2005.Magnetic, Rosales.2021.Competition}. In that approach, LLM is tuned in discrete steps, each corresponding to a different sample. To study the phase transitions between WC and FQH at fixed filling, it would be important to realize both phases in the same sample and tune between them continuously. In particular, the evolution of observables (such as the gap) across the transition could determine its order and universal properties. 
\begin{figure*}
\centering
\includegraphics[width=1\textwidth]{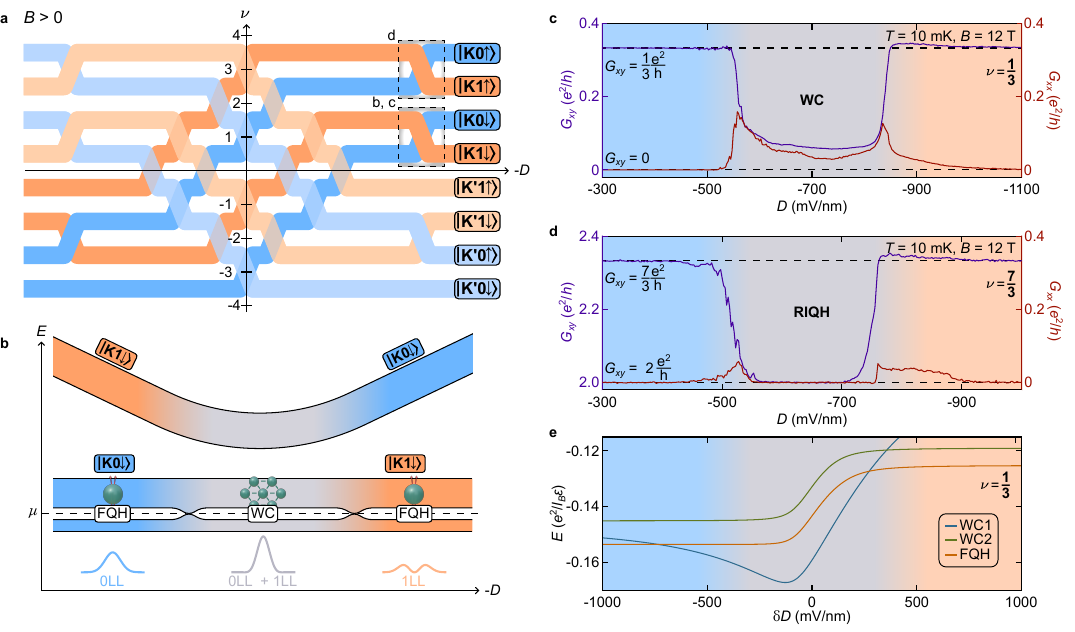}
\caption{\textbf{FQH--WC and FQH--RIQH competition at Landau-level crossings.} (\textbf{a}) Evolution of symmetry-broken LLs with displacement field. (\textbf{b}) Illustration of crossing $N=0$ and $N=1$ Landau levels as a function of $D$. (\textbf{c}) Transverse and longitudinal conductances, $G_{xy}$ and $G_{xx}$, as a function of $D$, show the transition from the $\nu=\frac{1}{3}$ plateau to an insulating state and back. (\textbf{d}) $G_{xy}$ and $G_{xx}$ as a function of $D$ showing the transition from the $\nu=\frac{7}{3}$ plateau to the RIQH effect and back. (\textbf{e}) Numerically obtained energies in units of Coulomb energy of competing Laughlin FQH and WC states as a function of $D$ relative to the crossing point $\delta D$. Here, $\varepsilon$ is the permittivity and $\ell_B$ is the magnetic length. For strong positive or negative $\delta D$, the FQH is energetically favorable in pure $N=0$ and $N=1$ levels, while the WC is favored near the crossing.}
\label{fig:figure1}
\end{figure*}
Graphene-based van der Waals heterostructures offer a natural arena for probing this physics. Their electrostatically controlled carrier density could realize the transition in a single device. BLG offers additional tunability due to its Landau-level spectrum's dependence on displacement field. In BLG at a fixed out-of-plane magnetic field, applying a displacement field $D$ lifts the valley degeneracy of the spin and orbital polarized LLs, generating various LL crossings within the eightfold zeroth LL of BLG(Fig.~\ref{fig:figure1}a). At moderate $D$, the phase diagram exhibits sixteen crossings, which have been studied previously \cite{hunt2017direct, Li.2017.Even, Zibrov.2017.Tunable, Huang.2022.Valley,Kumar.Quarter.2025}. At much larger $|D|$, additional, recently observed crossings emerge \cite{xiang2023intra, Lambert_ferromagnetic_2013, Shizuya_Orbital_2021, Pan.2017.Layer}. These involve LLs with identical spin and valley quantum numbers but different orbital indices, a configuration that can enhance LLM.

In this study, we take advantage of those large-$D$ crossings, observing FQH--WC transitions tuned via the displacement field $D$ at many filling factors. The WCs occur over a range near a crossing of two Landau levels with equal spin and valley but different orbital quantum numbers, $N=0$ and $N=1$. They are flanked by FQH states on either side of the crossing. We observe that the activation energies of FQH states and WCs at fillings such as $\nu=\frac{1}{3},\frac{2}{5},\frac{7}{3}$ vanish continuously at the transition occurring at smaller $|D|$ and evolve more sharply at the higher $|D|$ transitions. Remarkably, at half fillings, we observe the formation of incompressible states as intermediate phases between the composite Fermi liquids and the WCs.

\section*{Competition between FQH liquids and electronic crystals}
Our two devices (D1 and D2) consist of hBN-encapsulated BLG with top and bottom graphite gates, allowing for the independent control of $\nu$ and $D$ (see appendix A). After fabrication into Hall bars, the devices were measured in a top-loading dilution refrigerator with a base temperature of $T=10~\mathrm{mK}$ and a magnetic field of up to $B=12~\mathrm{T}$. The data shown in the manuscript are primarily from D1 and are consistent with those of D2; see Supplemental Material (SM) data at \cite{SM}.

The essence of our results is captured in Fig.~\ref{fig:figure1}. At very large displacement fields ($|D| \gtrsim 500~\mathrm{mV/nm}$ at $12~\mathrm{T}$) lead to an inversion of the $N=0$ and $N=1$ levels with the same spin and valley quantum numbers \cite{xiang2023intra}. The $N=0$ levels fully reside on one graphene layer, and thus, their energies are affected more strongly by the displacement field rather than the energies of the $N=1$ levels, which have non-zero weights on both graphene layers. At $D=0$, the $N=0$ levels have lower energy than $N=1$ with the same spin and valley quantum numbers, hence the electron-doped levels  $(\nu >0 )$ cross at large $|D|$ ($\gtrsim 500~\mathrm{mV/nm}$ at $12~\mathrm{T}$)~\cite{xiang2023intra}. At fixed fractional filling factors $0<\nu<1$ and $2<\nu<3$, the character of the partially filled level changes from $N=0$ at low $D$ to $N=1$ at larger $D$ through an avoided crossing, as illustrated in Fig.~\ref{fig:figure1}b ; see also SM~\ref{app.numerics} for the calculation of Landau levels energies as function of displacement field. At $\nu=\frac{1}{3}$ and $\nu=\frac{7}{3}$, FQH states form on either side of the crossings. Figs.~\ref{fig:figure1}c and \ref{fig:figure1}d show well-quantized fractional Hall conductances $G_{xy} =\frac{1}{3}$ and $G_{xy} =\frac{7}{3}$, respectively, in units of $e^2/h$, while the longitudinal conductances satisfy $G_{xx} \ll G_{xy}$.
Between these FQH regions, the transport properties change qualitatively, indicating phase transitions into a WC at $\nu=\frac{1}{3}$ and into a reentrant integer quantum Hall (RIQH) state at $\nu=\frac{7}{3}$; see Figs.~\ref{fig:figure1}c and \ref{fig:figure1}d. At $\nu=\frac{1}{3}$, $G_{xy}$ ceases to be quantized and decreases, while $G_{xx}$ increases and becomes comparable to $G_{xy}$, consistent with a weakly pinned WC \cite{Andrei.1988.Observation, Goldman.1990, Jiang.1990, Chen.2006.Melting, Deng.2016.Commensurability, maryenko2018composite, Rosales.2021.Competition, Singh.2024.Developing, Madathil.2023, Madathil.2024}. We attribute the non-zero values of $G_{xy}$ and $G_{xx}$ to geometrical properties of the device and the measurement parameters (see SM Sec.~1 at \cite{SM}). At $\nu=\frac{7}{3}$, the Hall conductance becomes integer, $G_{xy}=2$ with zero $G_{xx}$, as expected for a RIQH state. It can be regarded as a WC in the partially filled level on top of an integer quantum Hall effect from the fully filled levels \cite{Chen.2019.Competing, Liu.2012, Liu.2014, Shingla.2021}. Near the transitions between FQH and WC or RIQH behavior, the longitudinal conductance exhibits narrow peaks; for data from D2, see SM Sec.~2 at \cite{SM}.

The physical origin of the WC is illustrated in Fig.~\ref{fig:figure1}b. By forming superpositions of $N=0$ and $N=1$ orbitals, electrons can form more narrowly localized wave functions, which are favorable to Wigner crystallization. Fig.~\ref{fig:figure1}e shows a Hartree--Fock calculation comparing the energies of two WCs to the $\nu=\frac{1}{3}$ FQH state for unscreened Coulomb interactions. We consider two distinct WC trial states that are favored in different regimes. The first WC (WC1) incorporates the narrowly localized wave functions by mixing states with the same angular momentum and is favored at the Landau-level crossing, while FQH states have lower energies on either side of it. The second WC (WC2) is formed of superpositions of states with the smallest angular momentum and is favored over WC1 in a pure $N=1$ level, where the lowest angular momentum state of WC2 is more localized than the zeroth angular momentum state of WC1; see Appendix Sec.~\ref{SM:numerics} for details. For small deviations of the displacement field $\delta D = D-D_0$ from the level crossing point $D_0$, the WC1 is energetically favored; for large $|\delta D|$, the Laughlin state has the lowest energy. The numerical results shown in Fig.~\ref{fig:figure1}e capture the transition qualitatively, but find WC1 being favorable over a wider range than is observed experimentally. This discrepancy arises most likely because the energy of the $\nu=\frac{1}{3}$ state is overestimated by the Laughlin wavefunction, which best captures the ground state of $N=0$ without LLM. For a further discussion of this point and additional numerics using screened Coulomb interactions, see SM~Sec.~\ref{app.numerics}

\section*{FQH--WC--FQH transition at \texorpdfstring{$\nu=1/3$}{nu = 1/3}}
\begin{figure*}
 \centering
 \includegraphics[width=\textwidth]{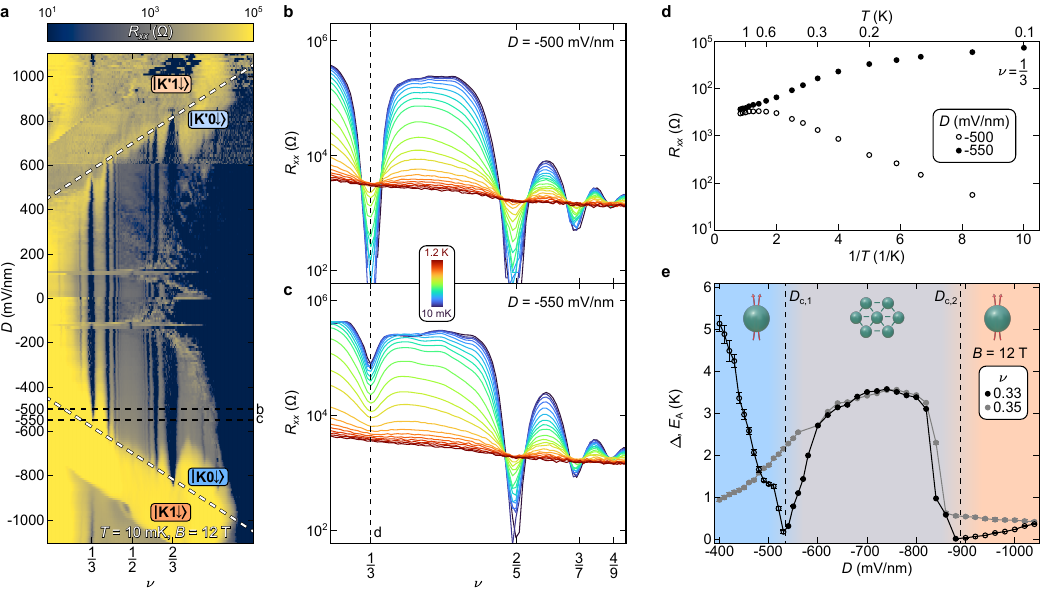}
 \caption{\textbf{FQH--WC--FQH phase transitions.} (\textbf{a}) $R_{xx}$ as a function of $\nu$ and $D$ showing the disappearing and reemerging FQH states. Black dashed lines indicate data shown in panels b and c. The white dashed lines indicate the expected crossing of partially filled $N=0$ and $N=1$ levels. Colored labels mark quantum numbers of Landau levels. Crossings at lower $|D|$ were omitted for clarity. (\textbf{b}) and (\textbf{c}) $R_{xx}$ as a function of $\nu$ for different temperatures at $D=-500~\mathrm{mV/nm}$ and $D=-550~\mathrm{mV/nm}$, respectively. (\textbf{d}) Arrhenius plot of $R_{xx}$ highlighting the opposite temperature dependence of $R_{xx}$ above and below $D_{c,1}$. (\textbf{e}) Activation energies of FQH and WC states at $\nu=\frac{1}{3}$ showing the FQH--WC--FQH transition. Closed circles were obtained by fitting to $R_{xx} \propto \exp\left(\frac{E_\mathrm{A}}{2k_\mathrm{B}T}\right)$, open circles with $R_{xx} \propto \exp\left(-\frac{\Delta}{2k_\mathrm{B}T}\right)$.}
 \label{fig:figure2}
\end{figure*}

Fig. \ref{fig:figure2}a shows $R_{xx}$ as a function of filling factor and displacement field at $B = 12~\mathrm{T}$ and $T = 10~\mathrm{mK}$. At low $|D|$, we observe a myriad of Jain states that are replaced by half-filled states in two pockets around $D \approx \pm 120~\mathrm{mV/nm}$. Additionally, quarter-filled states have been observed in the same device at larger $B$ \cite{Kumar.Quarter.2025}, attesting to the high quality of the device. The FQH states at low $|D|$ are separated by resistive (large-$R_{xx}$) regions. These regions smoothly connect to a resistive strip extending over \textit{all} fillings at $|D| \approx 800~\mathrm{mV/nm}$. The resistive strip follows the expected crossing of partially filled $N=0$ and $N=1$ levels, indicated by the white dashed line whose slope reflects the different Coulomb energies of uncorrelated electronic liquids in $N=0,1$; see also Appendix Sec.~\ref{SM:numerics}. The onset of the resistive region at the lower $|D|$ boundary is relatively gradual, in contrast to a much sharper higher-$|D|$ boundary. Positive and negative displacement fields show similar behavior; we focus on negative $D$, where the electrostatic doping of the contacts is optimal. Fig.~\ref{fig:figure2} shows data from D1; see SM Sec.~4 at \cite{SM} for data from D2.

The longitudinal resistance at $\nu=\frac{1}{3},\frac{2}{5},\frac{3}{7}$ at $D=-500~\mathrm{mV/nm}$ increases with temperature as expected for FQH states; see Fig.~\ref{fig:figure2}b. In contrast, the resistance decreases between those fillings, which we interpret as the formation of WCs, consistent with earlier STM measurements at lower $|D|$ \cite{Tsui.2024}. Upon raising the displacement field toward the resistive region, the temperature dependence at $\nu=\frac{1}{3}$ reverses; see Fig.~\ref{fig:figure2}c and the Arrhenius plot in Fig. \ref{fig:figure2}d. See SM Sec.~8 at \cite{SM} for the corresponding $R_{xy}$. Similar reversals occur at other fillings above different displacement fields; see SM Sec.~7 at \cite{SM}. 2D plots of $R_{xx}$ as a function of $\nu$ and $D$ measured at different temperatures are shown in SM Sec.~14 at \cite{SM}. Based on these findings, along with non-linear $I$--$V$ characteristics (SM Sec.~10 at \cite{SM}), we attribute the resistive strip to a single WC extending over all filling factors. This interpretation suggests that transitions between FQH states and WCs occur along many trajectories of constant $\nu$.

Fig.~\ref{fig:figure2}e shows the activation energy extracted from the temperature dependence of $R_{xx}$ for $\nu=\frac{1}{3}$ at different displacement fields. Specifically, we fit $R_{xx} \propto e^{- \Delta/2k_\mathrm{B}T}$ when the resistance increases with temperature, above $D_{c,1}=-535~\mathrm{mV/nm}$ and below $D_{c,2}=-890~\mathrm{mV/nm}$. In between, the resistance decreases, and we use $R_{xx} \propto e^{+ E_\mathrm{A}/2k_\mathrm{B}T}$ \cite{Jiang.1990, Wang.2023.Fractional, Singh.2024.Developing, Wang.2025.Competing}. Starting in the FQH regime above $D_{c,1}$, the activation gap $\Delta$ gradually closes upon approaching the transition. On the other side of this transition, $E_\mathrm{A}$ characterizes the strength of the WC and sets on with a slope mirroring the FQH activation gap. In contrast, near the $D_{c,2}$ transition, $E_\mathrm{A}$ drops sharply and $\Delta$ opens very slowly.

In between $\nu=\frac{1}{3}$ and $\nu=\frac{2}{5}$, at $\nu=0.35$, the sample is always resistive and $E_\mathrm{A}$ gradually increases with $|D|$. Eventually, it merges with that at $\nu=\frac{1}{3}$. At $D_{c,2}$, $E_\mathrm{A}$ drops sharply, indicating a transition into another resistive state. The observed FQH--WC--FQH transition follows the increase of the LLM parameter $\kappa(D)  =E_\mathrm{C}/\Delta(D)$, as the $N=0$ and $N=1$ LLs approach. From activation gap measurements at the $\nu=1$ crossing, we obtain $\Delta\approx 10~\mathrm{K}$, corresponding to $\kappa\approx 60$, substantially larger than typical values in other platforms ($\kappa\approx10$ \cite{Rosales.2021.Competition}; SM Sec.~11 at \cite{SM}). For a discussion of $R_{xx}$ as a function of $1 \leq \nu \leq 2$ and $D$, see SM Sec.~13 at \cite{SM}.

\begin{figure*}
 \centering
 \includegraphics[width=\textwidth]{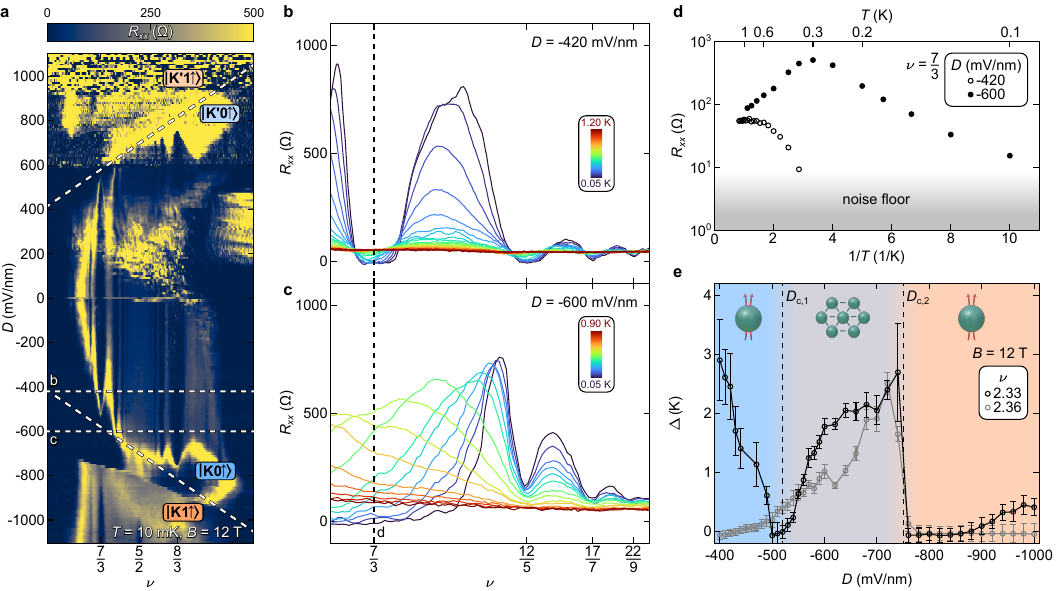}
 \caption{\textbf{FQH--RIQH--FQH phase transitions at $2<\nu<3$.} (\textbf{a}) $R_{xx}$ as a function of $\nu$ and $D$ showing the disappearing and reemerging FQH effect. Colored labels mark quantum numbers of Landau levels. Crossings at lower $|D|$ were omitted for clarity. (\textbf{b}) and (\textbf{c}) $R_{xx}$ as a function of $\nu$ for different temperatures at $D=-420~\mathrm{mV/nm}$ and $D=-600~\mathrm{mV/nm}$, respectively. (\textbf{d}) Arrhenius plot of $R_{xx}$ showing similar temperature dependence of $R_{xx}$ above and below $D_{c,1}$. (\textbf{e}) Activation gaps of FQH and RIQH states across the two transitions. All gaps were obtained by fitting $R_{xx}\propto\exp\left(-\frac{\Delta}{2k_\mathrm{B}T}\right)$.} 
 \label{fig:figure3}
\end{figure*}
\section*{FQH--RIQH--FQH transition at \texorpdfstring{$\nu=7/3$}{nu = 7/3}}
Fig.~\ref{fig:figure3} repeats the $R_{xx}$ measurements shown in Fig.~\ref{fig:figure2} in the partially filled Landau level $2<\nu<3$. Oppositely to the previous case, the strip separating the $N=0$ FQH states at moderate $|D|$ from the $N=1$ FQH states at large $|D|$ has $R_{xx}=0$. This difference arises due to the presence of two filled Landau levels contributing $2 \frac{e^2}{h}$ to the Hall conductance. Consequently, $R_{xy}=\frac{1}{2} \frac{h}{e^2}$ and $R_{xx}=0$ when the partially filled level realizes a WC---the RIQH effect \cite{Shingla.2021, Myers.2021, Chen.2019.Competing}.

Figs.~\ref{fig:figure3}b and \ref{fig:figure3}c show the temperature dependence of $R_{xx}$ as a function of $\nu$ for $D=-420~\mathrm{mV/nm}$ and $D=-600~\mathrm{mV/nm}$, respectively (see SM Sec.~9 at \cite{SM} for  $R_{xy}$). At $\nu=\frac{7}{3}$, Figs.~\ref{fig:figure3}b and \ref{fig:figure3}c correspond to the FQH and RIQH states, respectively, both showing $R_{xx}$ increasing with temperature; see also the Arrhenius plot in Fig.~\ref{fig:figure3}d. The activation gaps as a function of the displacement field are shown in Fig.~\ref{fig:figure3}e. At the $D_{c,1}=-520~\mathrm{mV/nm}$ transition, the gap closes symmetrically from both sides. At the $D_{c,2}=-750~\mathrm{mV/nm}$ transition, the RIQH gaps drop abruptly, and the FQH sets on very slowly, similarly to the behavior at $\nu=\frac{1}{3}$. The emergence of a gap at $\nu = \frac{5}{2}$ at $D < -1000~\mathrm{mV/nm}$ confirms a transition to an orbital index of $N=1$, a prerequisite for the emergence of half-filled states in BLG \cite{Li.2017.Even, Zibrov.2017.Tunable, Huang.2022.Valley, Kumar.Quarter.2025}. For a discussion of $R_{xx}$ as a function of $3 \leq \nu \leq 4$ and $D$, see SM Sec.~13 at \cite{SM}. For the hole side, see SM Sec.~16 at \cite{SM}.


\section*{Half-filled states at Landau-level crossings}
\begin{figure}
 \centering
 \includegraphics[scale=0.95]{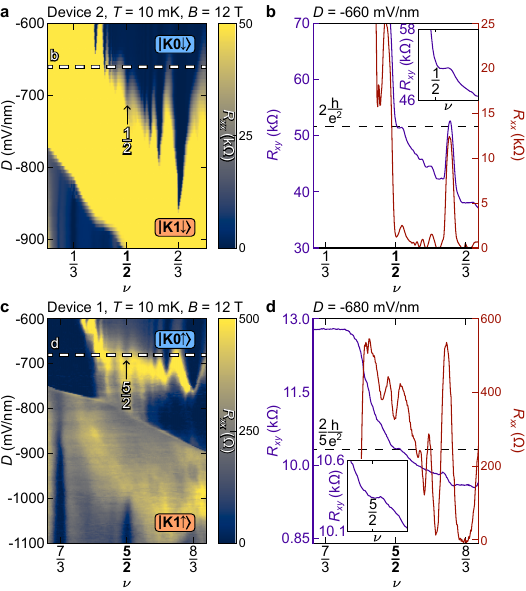}
 \caption{\textbf{Half-filled states at Landau-level crossings.} (\textbf{a}) $R_{xx}$ as a function of $\nu$ and $D$ centered around $\nu=\frac{1}{2}$. The $R_{xx}$ minimum at $\nu=\frac{1}{2}$ is marked by a black arrow. (\textbf{b}) $R_{xx}$ and $R_{xy}$ as a function of $\nu$ at $D=-660~\mathrm{mV/nm}$, clearly showing a minimum and a plateau, respectively. The dashed line marks the expected $R_{xy}$ value for $\nu=\frac{1}{2}$. (\textbf{c}) $R_{xx}$ as a function of $\nu$ and $D$ centered around $\nu=\frac{5}{2}$. The $R_{xx}$ minimum at $\nu=\frac{5}{2}$ is marked by a black arrow. (\textbf{d}) $R_{xx}$ and $R_{xy}$ as a function of $\nu$ at $D=-680~\mathrm{mV/nm}$, clearly showing a minimum and a plateau, respectively. The dashed line marks the expected $R_{xy}$ value for $\nu=\frac{5}{2}$.}
 \label{fig:figure4}
\end{figure}

Lastly, we turn our attention to the half-filled Landau levels at $\nu=\frac{1}{2}$ and $\nu=\frac{5}{2}$. Specifically, Figs.~\ref{fig:figure2}a and \ref{fig:figure3}a show faint features at these fillings just above the WC or RIQH regions. Fig.~\ref{fig:figure4}a shows a higher-resolution $R_{xx}$ map focusing on $\nu=\frac{1}{2}$ close to the FQH--WC transition. The dashed line indicates a cut along which $R_{xx}$ and $R_{xy}$ are plotted in Fig.~\ref{fig:figure4}b. At $\nu=\frac{1}{2}$, $R_{xx}$ develops a minimum, accompanied by a clearly resolved plateau in $R_{xy}$. At $\nu=\frac{5}{2}$, there is also a clear minimum in $R_{xx}$, but the accompanying plateau is less well developed; see SM Sec.~6 at \cite{SM}. Fig.~\ref{fig:figure4}c, therefore, shows an $R_{xx}$ map from the second device (D2) and Fig.~\ref{fig:figure4}d the corresponding linecuts which feature a clear plateau in $R_{xy}$. The emergence of these half-filled states can be understood as the mixing of $N=0$ with $N=1$ orbitals, which favor pairing. This picture is consistent with numerical studies supporting pairing in $N=0$ levels when LLM is included perturbatively \cite{Zhao_Composite_2023}. The resulting ground state is likely a non-Abelian 221 parton state \cite{Jain_Incompressible_1989, Wu_Non-Abelian_2017, Bandyopadhyay_Entangled_2018}.

\section*{Discussion}
The crossing of $N=0$ and $N=1$ levels with identical spin and valley is a unique feature of BLG at a large displacement field. Tuning across this level crossing allows continuous control over the LLM, enabling us to observe two manifestations of electronic crystals: The WC for $0<\nu<1$ and the RIQH state for $2<\nu<3$. In particular, a single WC phase extends over the full range of fillings, indicating a WC of electrons and not of holes or composite fermions. At fillings supporting FQH plateaus at weaker displacement fields, we observe that the activation energy vanishes continuously upon approaching the WC or RIQH phase, e.g., see black data in Fig.~\ref{fig:figure3}e. On the other side of this phase transition, the energy scale of the electron crystal states mirrors this behavior. In between plateaus, the WCs extend smoothly to low displacement fields while the RIQH states transition into resistive states, as can be seen in gray data in Fig.~\ref{fig:figure3}e. Qualitatively different behavior occurs at the second transition at larger displacement fields. The energy scale of the electronic crystal drops sharply at all fillings, and the phase boundary evolves smoothly with $\nu$. Observing a sharp transition at generic fillings, where a gradual crossover from a WC to a diffusive state is expected, could indicate an intermediate phase, e.g., a competing crystalline state undergoing a first-order transition into the primary WC. However, despite the sharpness of the second transition, we did not observe any hysteresis in the transport coefficients; see SM Sec.~12 at \cite{SM}.

These measurements strongly suggest that electron--electron interactions are the dominant factor in the observed transition, but cannot rule out that disorder also plays a significant role. In particular, the transport between plateaus at low displacement fields is likely to be diffusive. At these fillings, there is no sharp transition into the Wigner crystal, but a smooth crossover, cf.~Fig.~\ref{fig:figure3}e. Consequently, disorder is likely to play a role for any $\nu, D$ and transport measurements cannot directly disentangle it from interaction effects. Still, the occurrence of the transitions at fixed electron density and sample temperature cannot readily be accounted for by disorder alone. The interaction-driven interpretation is further supported by our observation that WCs extend over all $0<\nu<1$, in stark contrast to semiconductor structures. This behavior is precisely what one expects for WCs stabilized by strong mixing with the nearby empty level near the Landau-level crossing, which can only benefit WCs of electrons but not of holes. Finally, the emergent displacement-field-dependent energy scale in our activated transport measurements also indicates an interaction-based mechanism. 

In addition to promoting electronic crystals, LLM also facilitates pairing of composite fermions at half filling. At $\nu=\frac{1}{2}$ and $\frac{5}{2}$, the composite-Fermi liquid at moderate displacement fields first transitions into a paired state with a clear FQH plateau, before transitioning into the WC or RIQH state. Even-denominator states at Landau-level crossings were previously observed in GaAs hole systems \cite{Liu_even-denom_2014}. There, two $N=0$ levels formed by light and heavy holes cross, in contrast to the $N=0$ and $N=1$ levels crossing in BLG. The nature of the pairing in either case is not understood and will be an interesting question for future studies.

Our work highlights important directions for future research into the interplay of topological and symmetry-breaking states. First, the structure of the electronic crystals at high $|D|$ should be established directly. Additionally, the nature of the first transition, which appears to be continuous in our measurements, was not conclusively established in this work. 
Its systematic study would provide crucial insights into the mechanism of phase transitions between topological and conventional orders, including that between Chern insulators and charge density wave states \cite{Patri_Extended_2024, Song_Phase_2024, Lu_Extended_2025}.

\begin{acknowledgments}
It is a pleasure to thank Liang Fu, Jainendra Jain and Efrat Shimshoni for illuminating discussions. \textbf{Funding:} R.K. acknowledges support from the Dean of the Faculty and the Clore Foundation. Y.R. acknowledges the support from the Quantum Science and Technology Program 2021, the Schwartz Reisman Collaborative Science Program, supported by the Gerald Schwartz and Heather Reisman Foundation, supported by a research grant from the Goldfield Family Charitable Trust, the Minerva Foundation with funding from the Federal German Ministry for Education and Research, and the European Research Council Starting Investigator Grant Anyons 101163917. D.F.M. acknowledges support from the Israel Science Foundation (ISF) under grant 2572/21 and from the Minerva Foundation with funding from the Federal German Ministry for Education and Research.
\end{acknowledgments}

\section*{Author Contributions}
R.K. fabricated the devices. K.W. and T.T. grew the hBN crystals. A.H. developed the measurement codes. A.H. and R.K. performed the measurements. A.H., R.K., T.F., M.Y., D.F.M., and Y.R. analyzed the measured data. M.Y. and D.F.M. developed the theoretical calculations. A.H., R.K., T.F., M.Y., D.F.M., and Y.R. authored the paper with input from all coauthors. Y.R. supervised the overall work done on the project. 
\section*{Competing interests}
The authors declare no competing interests.

\section*{Data availability}
The data that support the plots in the main text are available at Zenodo repository \cite{data_zenodo}. All other data that support the findings of this study are available from the corresponding author upon reasonable request.

\appendix
\section{Materials and Methods}\label{Methods}
The heterostructures used in the two studied devices were fabricated using the well-known dry-transfer stacking technique \cite{Pizzocchero.2016.Hot}. The flakes of hexagonal boron nitride (hBN), BLG, and few-layer graphite for the gates were exfoliated on a silicon substrate with a $285\text{-}\mathrm{nm}$ SiO\textsubscript{2} layer. A polydimethylsiloxane stamp covered with polycarbonate (PC) was used to pick up the top graphite flake and then, subsequently, hBN, BLG, hBN, and bottom graphite flakes from the substrate. The heterostructure, along with the PC film, was then placed on a clean substrate of the same kind. The substrate was cleaned in chloroform for two hours to remove residual PC. The heterostructure was then vacuum-annealed at $400^\circ \mathrm{C}$ for three hours and an AFM tip in contact mode was used to iron the heterostructure \cite{kim2019reliable} and determine the thicknesses of individual flakes. The desired geometry was achieved by dry-etching (O\textsubscript{2} for graphite gates, CHF\textsubscript{3}/O\textsubscript{2} for hBN) \cite{wang_one_2013}. The BLG layer was then contacted using one-dimensional edge contacts \cite{wang_one_2013}. Special care was taken to improve the contact quality by optimizing the ratio of chromium ($2\text{--}3~\mathrm{nm}$), palladium ($13~\mathrm{nm}$), and gold ($70~\mathrm{nm}$) and depositing them at an angle of $\approx 15^{\circ}$ while continuously rotating the sample, leading to a low contact resistance of $\approx150~\Omega\mu\mathrm{m}$. After deposition, a lift-off procedure was performed in acetone and IPA. The final device's optical image is shown in panel \textbf{a} of Fig.~S1 in SM Sec.~1 at \cite{SM}. Four-probe resistance measurements were performed using the standard lock-in technique at $3.11~\mathrm{Hz}$ frequency. The measurement scheme is shown in Fig.~S1b in SM Sec.~1 at \cite{SM}. The measurement setup is placed in a shielded room and all ground loops were eliminated, thus reducing the electrical noise level.

All measurements were performed in the $n$--$D$ phase space with the desired values converted to top and bottom gate voltages according to
\begin{align}
\begin{split}
 V_\mathrm{tg} &= \frac{d_\mathrm{hBN, top}}{\varepsilon_\mathrm{hBN}} \left(\frac{n e}{2\varepsilon_0} - D\right) + V_\mathrm{CNP, top},\\
 V_\mathrm{bg} &= \frac{d_\mathrm{hBN, bottom}}{\varepsilon_\mathrm{hBN}} \left(\frac{n e}{2\varepsilon_0} + D\right) + V_\mathrm{CNP, bottom},
 \end{split}
\end{align}
where $d_\mathrm{hBN,(top,bottom)}$ are the top and bottom hBN thicknesses, $\varepsilon_\mathrm{hBN} = 3.9$ is the dielectric constant of hBN, and $V_\mathrm{CNP,(top, bottom)}$ is the top and bottom gate voltage corresponding to the charge-neutrality point of the device.

The device geometry reveals the origin of meandering artifacts in our $R_{xx}$ data. Due to the placement of the different gates, we obtain several interfaces at the edge: From the actual Cr/Pd/Au contact to a silicon-gated region on to another region that is gated by the graphite top gate and the silicon gate, and then the actual device. At negative (positive) $D$ on the hole (electron) side, the silicon bottom gate is at a negative (positive) voltage, and the graphite top gate is at a positive (negative) voltage, thus depleting the contact region of charge carriers, resulting in large contact resistance and, eventually, the artifacts seen in our $R_{xx}$ data.

\section*{Appendix B: BLG single particle Hamiltonian}\label{SM:numerics}
\subsection{Four-band model}
The transition from FQH to WC occurs when two BLG Landau levels, $N=0$ and $N=1$, with the same spin and valley approach each other as the displacement field $D$ changes. To describe the evolution of these levels, we closely follow Refs.~\cite{McCann_BLG_2006, McCann_BLG_2013, Jung_Accurate_BLG_2014, hunt2017direct} and use the values of parameters from Ref.~\cite{hunt2017direct}. In the basis $\Psi=(\psi_A,\psi_{B'},\psi_{B},\psi_{A'})$, the BLG Hamiltonian in the $K$-valley is given by 
\begin{align}\label{eq.H1BLG}
 H_{1}= \omega_0 \begin{pmatrix}
 \frac{u}{2 \omega_0} & 0 & a^\dag & \frac{\gamma_4}{\gamma_0}a^\dag\\
 0 & -\frac{u}{2 \omega_0}& \frac{\gamma_4}{\gamma_0}a & a 
 \\
 a & \frac{\gamma_4}{\gamma_0}a^\dag & \frac{u}{2 \omega_0} +\frac{\Delta}{\omega_0} & \frac{\gamma_1}{\omega_0}\\
 \frac{\gamma_4}{\gamma_0}a &a^\dag & \frac{\gamma_1}{\omega_0} & -\frac{u}{2 \omega_0} + \frac{\Delta}{\omega_0} 
 \end{pmatrix},
\end{align}
with $u$ the on-site energy difference between the two layers controlled by displacement field, and $\gamma_0=-2.61~\mathrm{eV}$, $\gamma_1=0.361~\mathrm{eV}$, $\gamma_4=0.138~\mathrm{eV}$, $\Delta=0.015~\mathrm{eV}$, and $\omega_0=\sqrt{\frac{3}{2}}\frac{a_0}{\ell_B}\gamma_0\approx 0.0306 \sqrt{B[\mathrm{T}]}~\mathrm{eV}$, where the distance between the two layers is $a_0=0.246~\mathrm{nm}$, while the Coulomb energy scale is $E_\mathrm{c}=\frac{e^2}{4\pi \epsilon_{\parallel} \ell_B}\approx 8.58 \sqrt{B[\mathrm{T}]}~\mathrm{meV}$. We take the dielectric constant as $\epsilon_\perp=3\epsilon_0$~\cite{hunt2017direct}, resulting in $u\approx 0.082~\mathrm{meV} \times D[\frac{\mathrm{mV}}{\mathrm{nm}}]$. 

The eigenvectors of $H_1$ spanning the zeroth Landau level are of the form
\begin{align} 
|\Psi_{N=0}\rangle &= (|0\rangle,0,0,0), \nonumber \\ |\Psi_{N=1}\rangle &= (C^1_A|1\rangle,0,C^0_{B}|0\rangle, C^0_{A'} |0\rangle) \label{eqn.sm.wfform} 
\end{align}
where $|n\rangle =[a^\dag]^n/\sqrt{n!}|0\rangle$ are the $n$th Landau level orbitals of non-relativistic particles. We restrict the Hamiltonian to the space spanned by $|\Psi \rangle =(C^1_A|1\rangle+C^0_A|0\rangle,0,C^0_{B}|0\rangle, C^0_{A'} |0\rangle)$. Then, the effective Hamiltonian acting on the vector $(C^0_A,C^1_A,C^0_{B}, C^0_{A'})$ is
\begin{align}
 H_\mathrm{ZLL} = \begin{pmatrix}
 \frac{u}{2 } & 0 & 0 & 0\\
 0 & \frac{u}{2 }& \omega_0 & \omega_0 \frac{\gamma_4}{\gamma_0} 
 \\
 0 & \omega_0 & \frac{u}{2 } +\Delta & \gamma_1\\
 0 & \omega_0 \frac{\gamma_4}{\gamma_0} & \gamma_1 & -\frac{u}{2 } +\Delta
 \end{pmatrix}.
\end{align}
The eigenvalue corresponding to $|\Psi_{N=0}\rangle$ is $E_{N=0} = \frac{u}{2}$. 
The spectrum of the lower $3\times3$-block has two solutions with energies $|\epsilon|>\gamma_1\gg\omega_0$ and one, $E_{N=1}$, that is close to $E_{N=0}$~\cite{Shizuya_Lamb_2012}. In the experimentally relevant regime where $\gamma_0,\gamma_1$ are much greater than other energy scales, we obtain the approximate single-particle energy of the $N=1$ state as
\begin{align}
 E_{N=1} = \frac{u}{2} - \eta u + \eta \left(\Delta - 2\frac{\gamma_1\gamma_4}{\gamma_0}\right),
\end{align}
where $\eta =\frac{\omega_0^2}{\gamma_1^2}\approx 7.2* 10^{-3} B[\mathrm{T}]$. 

\begin{figure}
 \centering
 \includegraphics[width=1.0\linewidth]{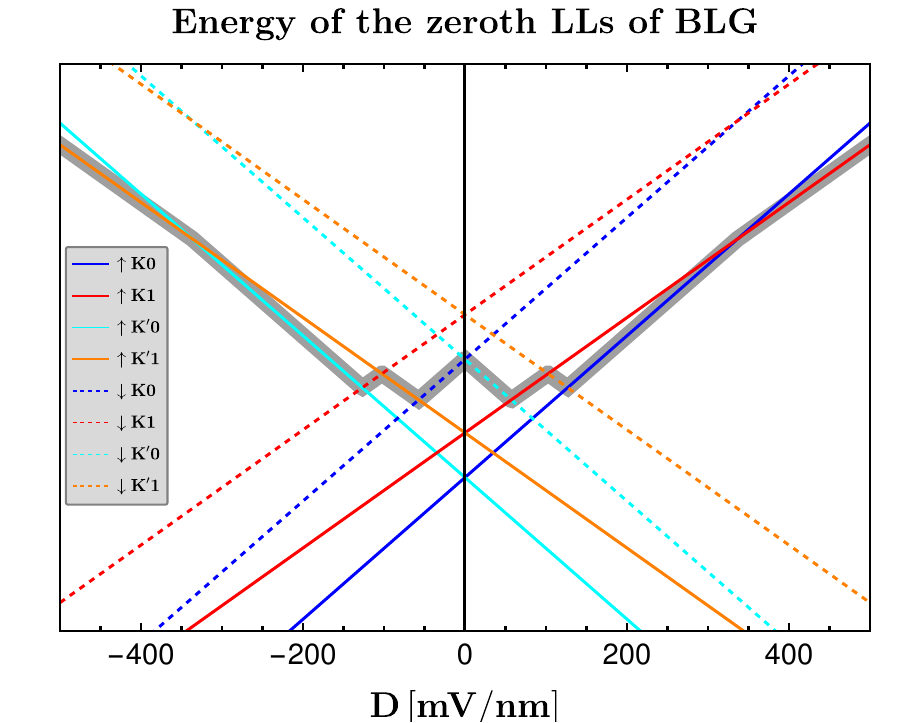} 
 \caption{The energies of the eight Landau levels of bilayer graphene are plotted as a function of $D$. The level crossings of the $N=0$ and $N=1$ Landau levels with the same spin and valley occur at $|D|\approx 343 mV/nm$. The partially occupied Landau level in the range $0<\nu<1$ is marked in gray for reference.}
 \label{fig.energy_blg_ll}
\end{figure}

The energy difference between the $N=0$ and $N=1$ levels is further offset by the Lamb shift~\cite{Shizuya_Lamb_2012}. We take the value from Ref.~\cite{hunt2017direct}, i.e.,
\begin{align}
 \Delta_\text{Lamb} = -\frac{1.7\sqrt{B[\mathrm{T}]}}{1+5.98/\sqrt{B[\mathrm{T}]}}~\mathrm{meV}.
\end{align}
Adding this energy to $E_{N=1}$ and using the parameters specified above, we find that the $E_{N=1}$ and $E_{N=0}$ levels cross at $D_0=343 \frac{\mathrm{mV}}{\mathrm{nm}}$ for $B=12T$ consistent with Ref.~\cite{xiang2023intra} and the boundary of the resistive region in Fig.~\ref{fig:figure2}a in the main text. The energies of the Landau levels are plotted in Fig.~\ref{fig.energy_blg_ll}.

The $N=0$ wave functions reside on a single layer of BLG, while the $N=1$ wave functions generally have contributions from both; cf.~Eq.~\eqref{eqn.sm.wfform}. We thus express the spatial wave functions as two-component vectors, whose entries describe the two layers, i.e.,
\begin{align} \label{eq.BLG_WF} \Phi_{N=0,x_0}(\mathbf{r}) &= \begin{pmatrix} \phi_{0,x_0}(\mathbf{r}) \\ 0 \end{pmatrix}, \nonumber \\ \Phi_{N=1,x_0}(\mathbf{r}) &= \begin{pmatrix} \cos(\theta)\phi_{1,x_0}(\mathbf{r}) \\ \sin(\theta)\phi_{0,x_0}(\mathbf{r}) \end{pmatrix}. \end{align}
Here $x_0$ is the guiding center coordinate, and $\phi_{n,x_0}$ is the standard $n$th LL wave function in the Landau gauge. The angle $\theta$ parametrizes layer polarization of the $N=1$ BLG LL as $P_1 \equiv \cos^2\theta-\sin^2\theta$. At weak magnetic and displacement fields, the polarization is approximately given by $P_1 \approx 1-2\eta \approx 1-0.014 B[\mathrm{T}]$\footnote{The sub-leading contribution to polarization in the large $\gamma_0,\gamma_1$ expansion enters as $P_1 \approx 1-2\eta + \eta\left(\eta +\frac{2\Delta}{\gamma_1} \left[\frac{\gamma_4}{\gamma_0} -\frac{\Delta}{\gamma_1} \right]\right) $ }. In the experimentally relevant regime $D\lesssim 10^3 \left[\frac{\mathrm{mV}}{\mathrm{nm}}\right]$, $\theta$ can be assumed to be constant and we take $P_1=0.84$, calculated from the four-band model at $B=12~\mathrm{T}$.

\subsection{Avoided level crossing}
We are concerned with the system's fate near the crossing of $N=0$ and $N=1$ levels, described by the effective Hamiltonian
\begin{align}
 H_1 = \sum_{x_0}\left( E_{N=1}c^\dag_{1,x_0}c_{1,x_0} + E_{N=0}c^\dag_{0,x_0}c_{0,x_0}\right)
\end{align}
where $c_{N,x_0}$ is the annihilation operator of a BLG electron in the Landau gauge. Experimentally, we observed that these levels experience an avoided crossing due to the term not present in the {\it ab initio} model Eq.~\eqref{eq.H1BLG}. Hence, we introduce the coupling $\tau$ between $N=0$ and $N=1$, whose value we infer experimentally to be $\tau=0.65-1.2~\mathrm{meV}$ in SM Sec.~11 at \cite{SM}. The final effective Hamiltonian reads

\begin{align} H_1 = \sum_{x_0} \Big[ &\epsilon (c^\dag_{0,x_0}c_{0,x_0} - c^\dag_{1,x_0}c_{1,x_0}) \nonumber \\ &+ \tau (c^\dag_{1,x_0}c_{0,x_0} +c^\dag_{0,x_0}c_{1,x_0}) \Big] \label{eq.H1_split} \end{align}
where $\epsilon = \frac{E_{N=0}-E_{N=1}}{2}$ depends linearly on the displacement field. Reducing the displacement field $D < D_0$, the system moves away from the transition. In the regime $\delta D\equiv D-D_0\ll 0$, the occupied level predominantly consists of $N=0$ BLG LL. In the opposite limit, $\delta D \gg 0$, the character of the occupied level changes to $|N=1\rangle$. 

\subsection{Density operator and form factors}
We expand the creation operator of electrons restricted to the $N=0,1$ states as $\psi^\dag(\vect{r}) = \sum_{N=0,1}\sum_{x_0} c^\dag_{N,x_0} \Phi_{N,x_0}(\vect{r})$, where $\Phi_{N,x_0}$ are given by Eq.~\eqref{eq.BLG_WF}. Then, the projected density operator is
\begin{align}\label{eq.Delta} \rho(\mathbf{r}) &= \int d^2r\; \psi^\dagger(\mathbf{r}) \psi(\mathbf{r})e^{i\mathbf{q}\cdot\mathbf{r}} \nonumber \\ &\equiv \sum_{N_1,N_2=0,1}{\cal F}_{N_1}^{N_2}(\mathbf{q}) \hat{\Delta}_{N_1,N_2}(\mathbf{q}), \\ \hat{\Delta}_{N_1,N_2}(\mathbf{q}) &= \sum_{x_0}c^\dagger_{N_1,x_0+\frac{q_y}{2}}c_{N_2,x_0-\frac{q_y }{2}}e^{i x_0q_x}. \end{align}
where we set the magnetic length to unity. The operator $\hat{\Delta}_{N_1,N_2}$ has the meaning of the projected guiding center coordinate density. The functions ${\cal F}^{N_2}_{N_1}(\vect{q})$ are BLG form factors and are given by 
\begin{align}
 {\cal F}^{N_2}_{N_1}(\mathbf{q}) = \int d^2r\; &\Phi^\dagger_{N_1}\left(x-\frac{q_y }{2},y\right) \nonumber \\
 &\cdot \Phi_{N_2}\left(x+\frac{q_y}{2},y\right) e^{ixq_x}.
 \label{eq.FormFactor_Integral}
\end{align}
In our case, we only need those with $N_1,N_2\in[0,1]$, which are given by
\begin{align}
 {\cal F}^{0}_{0} &= F_0^0, \quad {\cal F}^{1}_{0} = \cos(\theta)F^1_0, \nonumber \\
 {\cal F}^{1}_{1} &= \cos^2(\theta)F^1_1 + \sin^2(\theta)F^0_0,
 \label{eq.form_factors_split}
\end{align}
in terms of standard Landau level form factors~\cite{MacDonald_Introduction_1994} 
\begin{equation}
\begin{aligned}
 F^{n_2}_{n_1}(\mathbf{q}) = &\sqrt{\frac{n_2!}{n_1!}} \left( i \frac{q_x + iq_y}{\sqrt{2}} \right)^{n_1-n_2} \\
 &\times L_{n_2}^{n_1-n_2} \left( \frac{|\mathbf{q}|^2}{2} \right) e^{-\frac{|\mathbf{q}|^2}{4}}.
\end{aligned}
\label{eq.LandauFormFactor}
\end{equation}

\section{Hartree--Fock approximation}
The two-body interaction Hamiltonian is given by 
\begin{align}
 \hat{H} = \frac{1}{2}\int \frac{d^2q}{(2\pi)^2} \rho(\vect q)V(|\vect q|)\rho(-\vect q),
\end{align}
where $V(|\vect q) = \frac{2\pi}{\epsilon}\frac{1}{|\vect q|}$. We define the effective Hartree and Fock potentials as
\begin{align}
 V^\mathrm{H}_{N_1,N_2,N_1',N_2'}(\mathbf{q}) &= v(q){\cal F}_{N_1}^{N_2}(\mathbf{q}){\cal F}_{N_1'}^{N_2'}(-\mathbf{q}), \label{eq.Hartree} \\
 V^\mathrm{F}_{N_1,N_2,N_1',N_2'}(\mathbf{q}) &= \frac{1}{2\pi} \int d^2p\; V^\mathrm{H}_{N_1,N_2',N_1',N_2}(\mathbf{p}) \nonumber \\
 &\times e^{-i(p_x q_y - p_y q_x)}, \label{eq.Fock}
\end{align}

Then, within the Hartree--Fock approximation, the expectation value of the energy is
\begin{equation}
\label{eqn.supp.hf}
\begin{aligned}
 \langle \hat{H}\rangle = \frac{1}{2} \int \frac{d^2q}{(2\pi)^2} &\sum_{\{N_i\}} \Delta_{N_1,N_2}(\mathbf{q}) \\
 &\times V^\mathrm{HF}_{N_1,N_2,N_1',N_2'}(|\mathbf{q}|) \Delta_{N_1',N_2'}(-\mathbf{q}),
\end{aligned}
\end{equation}
where $ V^\mathrm{HF} \equiv V^\mathrm{H}- V^\mathrm{F}$ and $\Delta_{N_1,N_2}(\vect{q}) \equiv \langle\hat{\Delta}_{N_1,N_2}(\vect{q})\rangle$ is the expectation value of the guiding center density operator. The cohesive energy is then expressed as the energy per particle $E_\mathrm{coh} =\langle \hat{H}\rangle/N_\mathrm{e}$.

\subsection{Background energy}
To compare the energies in different Landau levels (LLs), we compute the total energy $E_\mathrm{tot}=E_\mathrm{coh} + U$, which includes the cohesive energy and the energy of an uncorrelated liquid at the same LL fillings. We consider Wigner crystals formed from the superposition of $N=0$ and $N=1$ states, i.e., from $|\Psi(\alpha)\rangle = \cos{\alpha}|\Phi_{N=0}\rangle + \sin{\alpha}|\Phi_{N=1}\rangle$.

The energy of the uncorrelated liquid at filling $\nu$ in the Landau level given by $|\Psi(\alpha)\rangle$ orbitals is
\begin{align} \label{eq.U}
 U_{\alpha} &= -\frac{\nu}{2}\int \frac{d^2q}{(2\pi)^2} V(|\mathbf{q}|)|{\cal F}_{\alpha}(\mathbf{q})|^2, \nonumber \\
 {\cal F}_{\alpha}(\mathbf{q}) &= \sum_{N_1,N_2} a_{N_1} a_{N_2} {\cal F}^{N_2}_{N_1}(\mathbf{q}),
\end{align}
where ${\cal F}_{\alpha}(\vect{q})$ are the form factors of $|\Psi(\alpha)\rangle$, and $a_{0} = \cos(\alpha)$ and $a_1=\sin(\alpha)$. For Coulomb interactions in units of $E_c$, we obtain 
\begin{align}
 U_{\alpha} &= -\frac{\nu}{2}\int_0^\infty \mathrm{d}q | {\cal F}_{\alpha}(\mathbf{q})|^2 \nonumber \\
 &= U_{N=0} + \sin^4\!\alpha (U_{N=1} - U_{N=0}),
 \label{eq.U_alpha_result}
\end{align}
where the last equality used results derived in Sec.~\ref{sm.formfactors} to express the energy in terms of the energies of Laughlin states in the $N$th BLG LL, i.e., 
\begin{align}
 U_{N=0} &= -\frac{\nu}{2}\sqrt{\frac{\pi}{2}}, \nonumber \\
 U_{N=1} &= -\frac{\nu}{8}\sqrt{\frac{\pi}{2}} \left[ 4\sin^2\!\theta + 3\cos^4\!\theta \right].
 \label{eq.U_values}
\end{align}
In Fig.~\ref{fig:figure2}a and Fig.~\ref{fig:figure3}a of the main text, we indicated the lines in the $\nu$--$D$ plane where the energies of $N=0$ and $N=1$ Landau levels filled with uncorrelated liquid at filling $\nu$ cross. They are obtained as the solutions $D(\nu)$ of 
\begin{align}
 E_{N=0}(D)+U_{N=0}(\nu) =
 E_{N=1}(D)+U_{N=1} (\nu),
\end{align}
where $U_{N}$ is evaluated for the screened Coulomb potential Eq.~\eqref{eq.Vscr}. 
\section{Trial states}
We now define the trial states for the FQH and WC phases in terms of $|\Psi(\alpha)\rangle$. For the WC phase, we consider electrons localized at the sites of a triangular lattice. The FQH liquid at $\nu=\frac{1}{3}$ is represented by the Laughlin wave function.

\subsection{FQH states in two Landau levels}
The Laughlin state at $\nu=\frac{1}{3}$ in the lowest LL is described by the wave function $\Psi_0(\{z_j\})=\prod_{j<k}(z_j-z_k)^{3}$. To obtain the corresponding $n$th Landau level wave function, we apply raising operators to all electrons, i.e., $\Psi_n(\{z_j\}) = \prod_{j}\frac{[a^\dag_{j}]^n}{\sqrt{n!}}\Psi_0(\{z_j\})$. This expression can be understood by considering the expansion of $\Psi_0$ in terms of 0LL orbitals 
\begin{align}
 \Psi_0(\{z_j\})=\prod_{j<k}(z_j-z_k)^{3} = \sum_{I} C_I \det[\phi_{0,m_I(j)}(z_i)]
\end{align}
and replacing $\phi_{0,m}\to \phi_{n,m}$ without altering the coefficients $C_I$. In our case, the Laughlin state is formed from the linear combinations of $N=0$ and $N=1$ BLG LL orbitals described by $|\Psi(\alpha)\rangle$. The interaction energy, derived in Eq.~\eqref{eq.rel2} below, is
\begin{align}
 E^\mathrm{int}_\mathrm{Laughlin}(\alpha) = \left[1-\sin^4\!\alpha\right] E_{N=0} + \sin^4\!\alpha E_{N=1},
\end{align}
expressed in terms of the energies $E_{N=0,1}$ in the pure $N=0$ and $N=1$ BLG LL. Using the appropriate Haldane pseudopotentials $H_m$, we compute the interacting energy from the static structure factors as outlined in Ref.~\cite{Goerbig_Competition_2004}. We find 
\begin{align}
 E_{N=0} &= \frac{\nu}{\pi} \sum_{m=0}^{3} c_{2m+1} H^{N=0}_{2m+1} + U_{N=0} \nonumber \\
 &\approx -0.4087, \label{eq.EN0} \\
 E_{N=1} &= \frac{\nu}{\pi}\sum_{m=0}^{3} c_{2m+1} H^{N=1}_{2m+1} + U_{N=1} \nonumber \\
 &\approx -0.3245 - 0.0975 \sin^2\!\theta + 0.0132\sin^4\!\theta. \label{eq.EN1}
\end{align}
where $(c_1,c_3,c_5,c_7)=(-1,17/32,1/16,-3/32)$.

The total energy of the $\nu=\frac{1}{3}$ Laughlin state constructed out of $|\Psi(\alpha)\rangle$ states is given by
\begin{equation} \label{eq.En_La}
\begin{split}
 E_{\text{Laughlin}}(\alpha) ={}& \epsilon(D) \cos(2\alpha) + \tau \sin(2\alpha) \\
 &+ [1-\sin^4\!\alpha] E_{N=0} + \sin^4\!\alpha E_{N=1}.
\end{split}
\end{equation}
Finally, we minimize the energy with respect to $\alpha$ for each displacement field, i.e.,
\begin{align}
 E_\text{Laughlin}(D) = \min\limits_{\alpha} \left[E_\text{Laughlin}(\alpha) \right].
\end{align}

\subsection{Inter-Landau-level coherent Wigner crystal}
We first define a WC trial state in a single $N=0$ BLG LL before introducing the inter-Landau-level coherent Wigner crystal (ILLC WC). Closely following Ref.~\cite{Fogler_Laughlin_wigner_1997}, we assume that the electron residing on a Wigner-lattice site at $\vect{R}_j$ forms a coherent state within the $N=0$ BLG LL. The guiding center coordinate operator $\vect{C}$ in the symmetric gauge is related to the intra-Landau level ladder operator $b$ via $\vect{C}=\left(\frac{b+b^\dag}{\sqrt{2}},i\frac{b-b^\dag}{\sqrt{2}}\right)$ and its components obey $[C_x,C_y]=i$. We define a coherent state centered around $\vect{R}_0=(X_0,Y_0)$ by
\begin{align}
 |\Phi_{N=0,\vect{R}_0}\rangle = e^{\frac{\bar{Z}_0b^\dag - Z_0 b}{\sqrt{2}}}|\Phi_{N=0,m=0}\rangle~,
\end{align}
where $Z_0=X_0+iY_0$ and $|\Phi_{N=0,m=0}\rangle$ is the $N=0$ BLG LL state in the symmetric gauge with zero angular momentum, such that $b|\Phi_{N=0,m=0}\rangle=0$. Here, $b$ is diagonal in the sublattice index, i.e., acts on both components of $\Phi_{N}$ in Eq.~\eqref{eq.BLG_WF}. The Wigner crystal wave function is obtained by arranging electrons into a triangular lattice 
\begin{align} \label{eq.WC}
 |\text{WC}_{N=0}\rangle &= \bigwedge_{j=1}^{N_e} |\Phi_{N=0,\mathbf{R}_j}\rangle, \nonumber \\
 \mathbf{R}_i &= j_1 \Lambda_0 \begin{pmatrix} 1 \\ 0 \end{pmatrix} + j_2 \Lambda_0 \begin{pmatrix} 1/2 \\ \sqrt{3}/2 \end{pmatrix}.
\end{align}
where $\Lambda_0=\sqrt{\frac{4\pi}{\sqrt{3}\nu}}$ is the lattice constant, and $\bigwedge_{j}$ is a wedge product over lattice sites. 

The interaction energy of the Wigner crystal decreases when the electronic states forming the lattice sites become more localized, i.e., have a smaller spread $\langle |\vect{R}-\vect{r}|\rangle$. The spread of the LL wave function in the symmetric gauge grows with the LL index $n$ and angular momentum $m$. This is why the typical WC trial state in $n$th LL is formed of coherent states of the smallest angular momentum state, i.e., $\phi_{n,-n}$. However, we note that the superposition $\cos(\alpha)\phi_{0,0} + \sin(\alpha)\phi_{1,0}$ of the 0LL and 1LL Landau level wave functions can have a smaller spread than either $\phi_{0,0}$, $\phi_{1,0}$, or $\phi_{1,-1}$; see Fig.~\ref{fig.alpha}a. For superpositions between non-relativistic LL wave functions, the spread is minimal at $\alpha_*=-\arctan(1/2)\approx-0.463$. For the BLG LL, where $N=1$ has both 0LL and 1LL contributions, the optimal $\alpha$ is different, and for the parameters used in our calculation, we find $\alpha_* \approx-0.502$. Hence, we expect the Wigner crystal formed of coherent superposition $N=0$ and $N=1$ states to be energetically favorable. 

Based on these considerations, we define the ILLC WC wave function 
\begin{align}
 |\text{WC}_\alpha\rangle = \bigwedge_{j=1}^{N_e} |\Psi_{\vect{R}_j}(\alpha)\rangle,
 \qquad
 |\Psi_{\vect{R}_j}(\alpha)\rangle = e^{\frac{\bar{Z}b^\dag - Zb}{\sqrt{2}}}
 |\Psi(\alpha)\rangle 
\end{align}
where $|\Psi(\alpha)\rangle =\cos(\alpha) |\Psi_{N=0,m=0}\rangle + \sin(\alpha) |\Psi_{N=1,m=0}\rangle$. The expectation value of the projected guiding center coordinate operator Eq.~\eqref{eq.Delta} 
\begin{align}\label{eq.coh}
 \Delta^{\alpha}_{N_1,N_2}(\mathbf{q}) &\equiv \langle\text{WC}_\alpha| \hat{\Delta}_{N_1,N_2}(\mathbf{q})|\text{WC}_\alpha\rangle \nonumber \\
 &= \begin{pmatrix}
 \cos^2\alpha {\cal F}_{0}^{0}(\bar{q}) & \cos\alpha\sin\alpha{\cal F}_{0}^{1}(\bar{q}) \\
 \cos\alpha\sin\alpha{\cal F}_{1}^{0}(\bar{q}) & \sin^2\alpha{\cal F}_{1}^{1}(\bar{q}) 
 \end{pmatrix}_{N_1,N_2} \nonumber \\
 &\quad \times \sum_{j} e^{i\mathbf{q}\cdot \mathbf{R}_j}
\end{align}
is computed as outlined in Section~\ref{sec.expectaion}. We compare the ILLC WC to an alternative trial state that superposes the $N=0$ and $N=1$ states with the smallest angular momenta $|\tilde\Psi(\alpha)\rangle =\cos(\alpha) |\Phi_{N=0,m=0}\rangle + \sin(\alpha) |\Phi_{N=1,m=-1}\rangle$ and obtain
\begin{align}\label{eq.Pcoh}
 \tilde{\Delta}^{\alpha}_{N_1,N_2}(\mathbf{q}) &\equiv \langle\tilde{\text{WC}}_\alpha| \hat{\Delta}_{N_1,N_2}(\mathbf{q})|\tilde{\text{WC}}_\alpha\rangle \nonumber \\
 &= F_{0}^{0}(\bar{q}) \begin{pmatrix}
 \cos^2\alpha & \frac{1}{2}\cos\theta\sin 2\alpha \\
 \frac{1}{2}\cos\theta\sin 2\alpha & \sin^2\alpha 
 \end{pmatrix}_{N_1,N_2} \nonumber \\
 &\qquad \times \sum_{j} e^{i\mathbf{q}\cdot \mathbf{R}_j}.
\end{align}

However, the interaction energy of this trial state does not benefit from mixing $N=0$ and $N=1$ orbitals. The two carry different angular momenta states, which prevents interference and does not permit a reduction in the wave function spread. 
Within the Hartree--Fock approximation of Eq.~\eqref{eqn.supp.hf}, the cohesive energy of the ILLC WC is 
\begin{equation} \label{eq.E_coh_WC}
\begin{split}
 E^\text{coh}_\text{WC}(\alpha) = \frac{\nu}{4\pi} \sum_{\mathbf{G} \neq 0} \sum_{\{N_i\}} &\Delta^{\alpha}_{N_1,N_2}(\mathbf{G}) \Delta^{\alpha}_{N'_1,N'_2}(-\mathbf{G}) \\
 &\times V^\text{HF}_{N_1,N_2,N'_1,N'_2}(\mathbf{G})
\end{split}
\end{equation}
where the sum is over all non-zero reciprocal lattice vectors $\vect{G}$. For the alternative trial state, one replaces $\Delta$ with $\tilde \Delta$. 

Finally, the total energy 
\begin{align}
 E_\text{WC}(\alpha) = \epsilon(D) \cos(2\alpha) + \tau \sin(2\alpha) + U_\alpha + E^\text{coh}_\text{WC}(\alpha)
\end{align}
is minimized with respect to $\alpha$ for each value of the displacement field 
\begin{align}
 E_\text{WC}(D) = \min\limits_{\alpha} \left[E_\text{WC}(\alpha)\right].
\end{align}

\section{Numerical results}\label{app.numerics}
\begin{figure}
 \centering
 \includegraphics[width=1.0\linewidth]{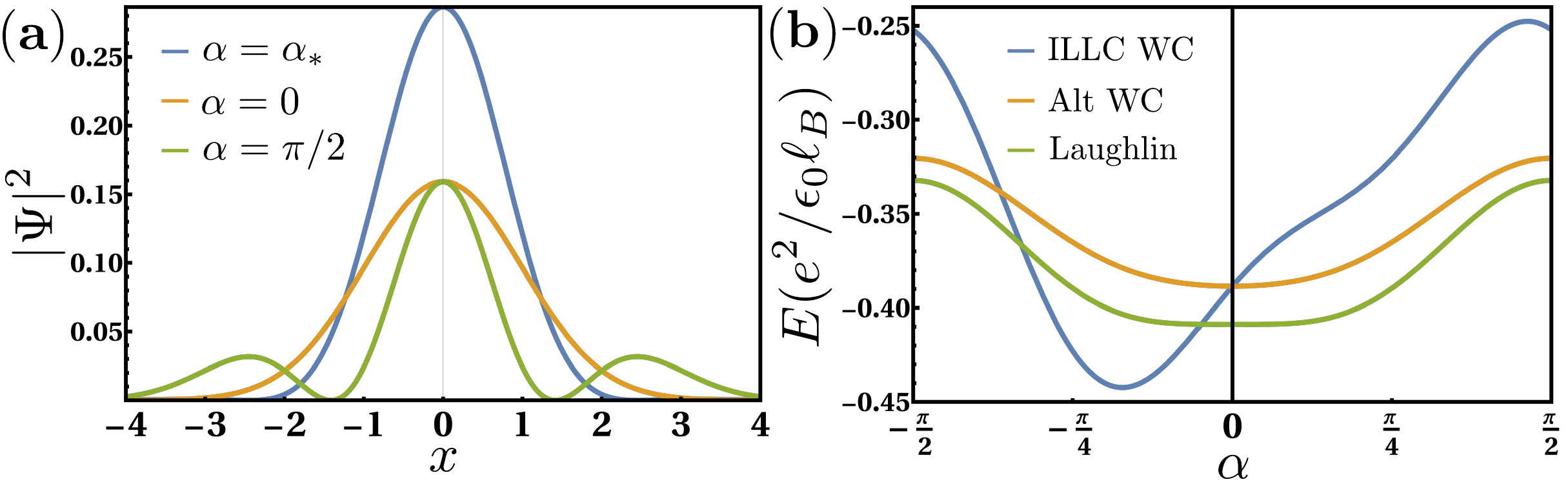} 
 \caption{The left panel shows the wave function spread as a function of mixing angle $\alpha$. For $\alpha=0,\pi/2$, pure 0LL and 1LL are wider than their coherent superposition with the optimal mixing angle $\alpha=\alpha_*=-\arctan(1/2)$ that minimizes the spread. The right panel shows the interaction energy of Laughlin Eq.~\eqref{eq.En_La}, ILLC WC Eq.~\eqref{eq.coh}, and alternative WC Eq.~\eqref{eq.Pcoh} trial states as a function of $\alpha$. The coherent Wigner crystal has the smallest energy at $\alpha\approx\alpha_*$, where the wave function has the smallest spread. }
 \label{fig.alpha}
\end{figure}

To understand the dependence of the trial energies of all three states on the variational parameter $\alpha$, we begin with the case of exactly degenerate $N=0$ and $N=1$ levels. Fig.~\ref{fig.alpha}b shows the energies given by Eq.~\eqref{eq.En_La}, Eq.~\eqref{eq.coh}, and Eq.~\eqref{eq.Pcoh} evaluated for $\epsilon=\tau=0$ and as a function of the mixing angle $\alpha$. The lowest energy of the Laughlin state and the alternative WC occur at $\alpha=0$, i.e., in a pure $N=0$ level. By contrast, the ILLC WC has a distinct minimum at $\alpha<0$, in agreement with the expectations based on wave function spread shown in Fig.~\ref{fig.alpha}b. Moreover, the minimal ILLC WC energy is below that of the Laughlin state, even though the Laughlin state is favored over a WC both in pure $N=0$ and pure $N=1$ levels.

\begin{figure}
 \centering
 \includegraphics[width=1.0\linewidth]{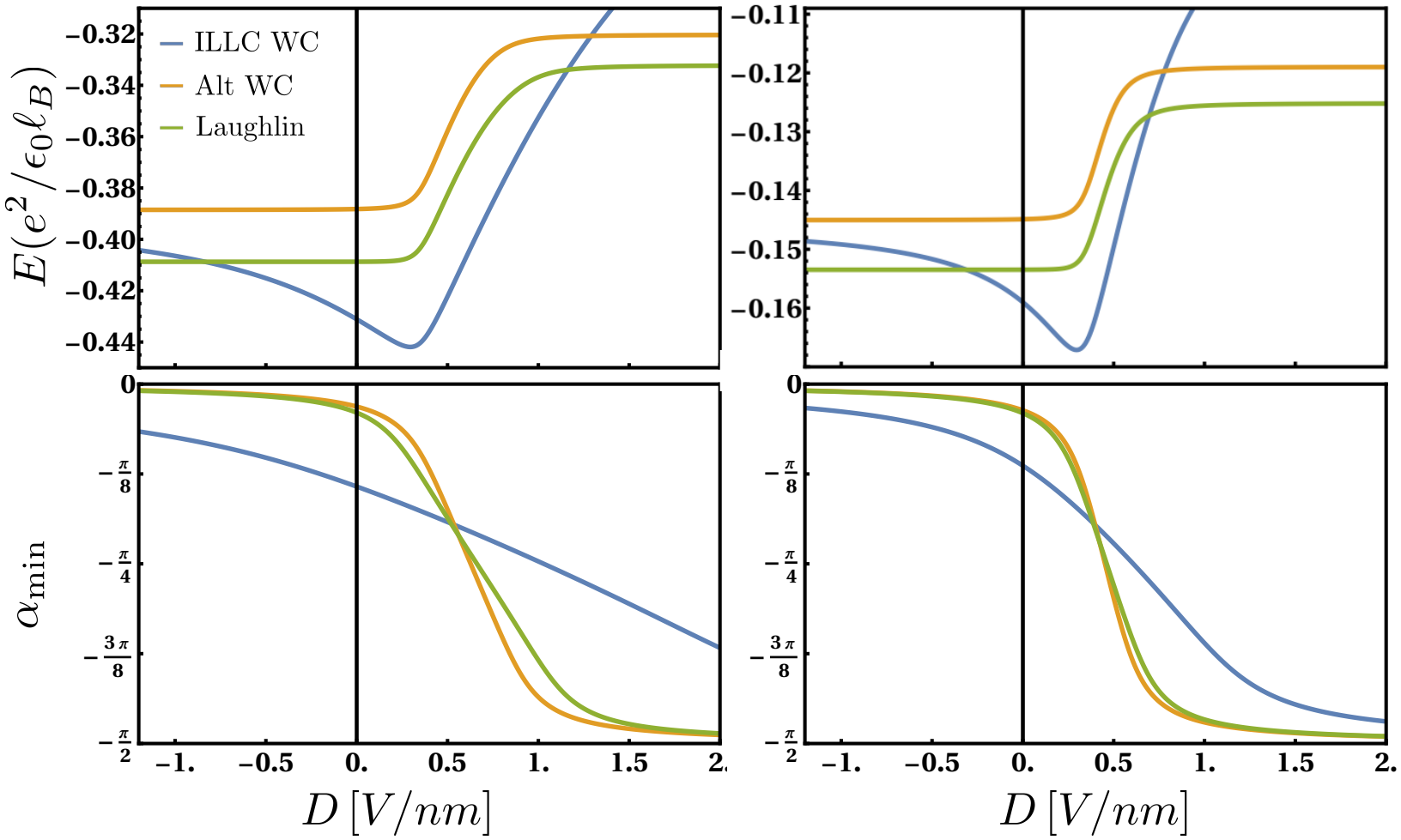} 
 \caption{The energies and the mixing angle $\alpha$ minimizing the energy is plotted for the range of the displacement fields $D$. In left panel, we use Coulomb interactions. In the right panel, we use the screened interactions Eq.~\eqref{eq.Vscr}. We note that transition occurs at the range of $D\sim [-0.5,0.7]\frac{\mathrm{V}}{\mathrm{nm}}$, while experimentally, we observe the WC favored over a much smaller range. We attribute this discrepancy to the fact that the Laughlin state does not benefit from occupying two Landau levels, while the exact ground state generally should have lower energy.}
 \label{fig.energy}
\end{figure}

Next, we restore the kinetic energy and minimize the trial energies of all three states for each value of the displacement field. The top left panel of Fig.~\ref{fig.energy} shows the optimal energies, which occur for the variational parameters plotted in the panel below. Near the crossing, the ILLC WC has the lowest energy, while Laughlin states are favored for sufficiently large $|D-D_0|\gg0$, consistent with the measurement.

This qualitative result is robust against changes in the microscopic form of the interactions. In the right panels of Fig.~\ref{fig.energy} we show the same quantities as on the left but for a screened interaction of the form
\begin{align}\label{eq.Vscr}
 V(q) &= \frac{V_0(q)}{1 + \frac{2\log 4}{\pi} a_\text{scr} V_0(q) \tanh(b_\text{RPA} q^2)}, \nonumber \\
 V_0(q) &= \frac{2\pi}{q} \tanh(q L_0).
\end{align}
We chose the parameters $a_\text{scr}=0.67 \frac{E_\mathrm{c}}{\omega_\mathrm{c}} = 2.19/\sqrt{B[\mathrm{T}]}$ and $b_\text{RPA}=0.62$ from Ref.~\cite{hunt2017direct} and the distance $L_0=20~\mathrm{nm}$ between metallic gates and BLG. The ILLC WC is again favored near $D_0$ but over a narrower region than for the pure Coulomb case.

Despite our calculations capturing the qualitative behavior of the transition correctly, we note that quantitative features are less well described. In particular, the predicted resistive region where ILLC WC is favored extends over a wider range of displacement fields for both interactions than measured experimentally. We attribute this discrepancy primarily to the fact that we are comparing the energies of specific trial states, which are not true ground states of the system. In particular, we expect the Laughlin wave function to capture the ground state energy very well in a pure $N=0$ level but less so for $N=1$ or near the LL crossing.

\section{Useful identities and relations}
\subsection{Form factors of isotropic interactions}\label{sm.formfactors}
For isotropic interactions $V(\vect{q})=V(|\vect{q}|)$, the calculation of various physically relevant quantities involves a momentum integral of the form factors $F_\psi(\vect{q})$ of the LL orbitals $\psi$ and a function that depends on the absolute value of momentum. For instance, for the energy of an uncorrelated liquid, we need to evaluate the integral
\begin{align}
 U_\psi &= \int \frac{\mathrm{d}^2q}{2\pi} |F(\mathbf{q})|^2 V(|\mathbf{q}|) \nonumber \\
 &= \int_0^\infty \frac{dq}{2\pi} qV(q) \int_0^{2\pi} \frac{\mathrm{d}\phi}{2\pi} |F_\psi(\mathbf{q})|^2.
 \label{eq.U_psi_integral}
\end{align}
Similarly, to determine Haldane pseudopotentials~\cite{Haldane_fqh_1983} $H_m$, we compute 
\begin{align}
 H_m=\int \frac{\mathrm{d}^2q}{2\pi} \;|F(\vect{q})|^2 V(|\vect q|) L_m(|\vect q|^2)e^{-|\vect q|^2/2}.
\end{align}
In both cases, the dependence on the direction of $\vect{q}$ enters only through $|F_\psi(\vect{q})|^2$. Hence, we can average $|F_\psi(\vect{q})|^2$ over $\phi\equiv \mathrm{arg}(\vect{q})$ without changing the result of the integral. When the orbitals $\psi$ are a combination of different Landau level orbitals $\phi_n$, i.e., $\psi=\sum_{n=0}^{r} a_n \phi_n$, it is possible to express the averaged form factors in terms of individual LL form factors
\begin{align}
 \int_0^{2\pi} \frac{\mathrm{d}\phi}{2\pi}\; |F_\psi(\vect{q})|^2 = \sum_{n=0}^r X^r_n(\{a_i\}) |F_n(\vect{q})|^2.
\end{align}
In particular, for $\psi=a_0\psi_0+a_1\psi_1$, we find
\begin{align}\label{eq.rel1}
 \int_0^{2\pi} \frac{\mathrm{d}\phi}{2\pi}\; |F_\psi(\vect{q})|^2 = [a_0^4+2a_0^2a_1^2]|F_0(\vect{q})|^2 + a_1^4|F_1(\vect{q})|^2.
\end{align}
This relation implies that the energy of any FQH states in $\psi = \cos(\alpha) \phi_0+\sin(\alpha)\phi_1$ LL orbitals obeys
\begin{align}\label{eq.rel2}
 E_\psi = \left(1-\sin^4\!\alpha\right)E_\text{0LL} + \sin^4\!\alpha E_\text{1LL},
\end{align}
where $E_{n\mathrm{LL}}$ are $n$th LL energies. The same relation holds for the energy of the uncorrelated liquid. For $\psi=a_0\phi_0+a_1\phi_1+a_2\phi_2$, we find
\begin{align}
 E_\psi &= \left[a_0^4 + a_1^2(\sqrt{2}a_0+a_2)^2\right] E_\mathrm{0LL} \nonumber \\
 &+ \big\{ [a_1^2+a_2^2][2a_0^2+a_1^2] \nonumber \\
 &\quad - 2a_0a_1^2[a_0+\sqrt{2}a_2] \big\} E_\mathrm{1LL} + a_2^4 E_\mathrm{2LL}.
 \label{eq.E_psi_LL}
\end{align}
Finally, we note that the relations provided in Eqs.~\eqref{eq.rel1} and \eqref{eq.rel2} hold for $N=0$ and $N=1$ BLG LL with the replacement $F\to{\cal F}$ .

\subsection{Evaluation of order parameters}\label{sec.expectaion}
It is convenient to compute the order parameter in the first quantized representation. For simplicity, we work with the standard Landau level; the generalization to BLG LL is straightforward. In symmetric gauge $\vect{A}=\frac{1}{2}(-y,x)$, the inter- and intra-LL ladder operators~\cite{Jain_composite_2007, MacDonald_Introduction_1994} are 
\begin{align}
 a &= \frac{1}{\sqrt{2}} \left[ \left(\frac{x}{2} + \partial_x\right) + i\left(\frac{y}{2} + \partial_y\right) \right], \nonumber \\
 b &= \frac{1}{\sqrt{2}} \left[ \left(\frac{x}{2} + \partial_x\right) - i\left(\frac{y}{2} + \partial_y\right) \right].
 \label{eq.LL_operators}
\end{align}
obeying $[a,a^\dag]=[b,b^\dag]=1$, and we use units where the magnetic length $\ell_B=1$. The density operator factorizes into inter- and intra-LL contributions
\begin{align}
 \rho(\vect{q})=\sum_{j}e^{i\vect{q}\cdot\vect{r}_j} = \sum_{j} e^{i\frac{\mathord{q} a^\dag_j + \bar{\mathord{q}} a_j}{\sqrt{2}}} e^{i\frac{\bar{\mathord{q}} b^\dag_j + \mathord{q} b_j}{\sqrt{2}}}~, 
\end{align}
where we defined complex momenta $\mathord{q}=q_x+iq_y$ and $\bar{\mathord{q}}=q_x-iq_y$. Introducing the guiding center coordinate operator $\vect{C}=\left(\frac{x}{2}- i \partial_{y}, \frac{y}{2}+i\partial_{x}\right)=\left(\frac{b+b^\dag}{\sqrt{2}},i\frac{b-b^\dag}{\sqrt{2}}\right)$, the last term is expressed
$e^{i\frac{\bar{\mathord{q}} b^\dag_j + \mathord{q} b_j}{\sqrt{2}}} =e^{i\vect{q}\cdot\vect{C}_j}$. Then, the projected density operator Eq.~\eqref{eq.Delta} in the first quantized from is defined as
\begin{align} \label{eq.order}
 \rho_{n_1,n_2}(\mathbf{q}) &\equiv \langle n_{1}| e^{i(q a^\dagger + \bar{q} a)/\sqrt{2}} |n_{2}\rangle \nonumber \\
 &\quad \times \left[ \sum_{j} |n_{1,j}\rangle e^{i\mathbf{q}\cdot \mathbf{C}_j} \langle n_{2,j}| \right] \nonumber \\
 &= F_{n_1}^{n_2}(\mathbf{q}) \times \Delta_{n_1,n_2}(\mathbf{q}).
\end{align}
The last equality implicitly defines the order parameter operator $\Delta_{n_1,n_2}(q)$, using the form factors \cite{MacDonald_Introduction_1994}
\begin{align}\label{eq.form}
 F_{n_1}^{n_2}(\mathbf{q}) &= \langle n_{1}| e^{i q a^\dagger / \sqrt{2}} e^{i \bar{q} a / \sqrt{2}} |n_{2}\rangle e^{-|q|^2/4} \nonumber \\
 &= \sqrt{\frac{n_2!}{n_1!}} \left(\frac{i q}{\sqrt{2}}\right)^{n_1-n_2} L_{n_2}^{n_1-n_2} \left(\frac{|\mathbf{q}|^2}{2}\right) e^{-\frac{|\mathbf{q}|^2}{4}}.
\end{align}

Closely following Ref.~\cite{Fogler_Laughlin_wigner_1997}, we define a coherent state in the $n$th Landau level localized at $\vect{R}_j$, by acting with the magnetic translation operator. Recall that guiding center coordinates obey canonical commutation relations $[C_x,C_y]=i$, and the coherent state of these operators, treated as a momentum and a coordinate, is
\begin{align}
 |\vect{R}_{0,0}\rangle = e^{\frac{\bar{Z}_0b^\dag - Z_0b}{\sqrt{2}}}|0,0\rangle, 
\end{align}
where $\vect{R}=(X_0,Y_0)$ and $Z_0=X_0+iY_0$, and $a|0,0\rangle=b|0,0\rangle=0$. This state is maximally localized around $\vect{R}$ in its guiding center coordinates. For the two WC trial states used in our calculations, we define the state that is obtained by acting with the magnetic translation operator $ e^{\frac{\bar{Z}_0b^\dag - Z_0b}{\sqrt{2}}}$ on the $m$th angular momentum state 
\begin{align}
 |\vect{R}_{0,m}\rangle = e^{\frac{\bar{Z}_0b^\dag - Z_0b}{\sqrt{2}}} |0,m\rangle = e^{\frac{\bar{Z}_0b^\dag - Z_0b}{\sqrt{2}}} \frac{[b^\dag]^{m}}{\sqrt{m!}}|0,0\rangle. 
\end{align}
This state is not an eigenstate of $b$. However, it is still well-localized around $\vect{R}$ for small $m$. The state in the $n$th Landau level is obtained by acting with raising operators 
\begin{align}
 |\vect{R}_{n,m-n}\rangle = \frac{[a^\dag]^n}{\sqrt{n!}} |R_{0,m}\rangle.
\end{align}
We now determine the matrix elements of $\hat\Delta$ between two coherent states. First, we note that all matrix elements vanish unless the Landau level indices coincide, i.e.,
\begin{align}
 \langle R_{n_1',m_1}|\hat{\Delta}_{n_1,n_2}|R_{n_2',m_2}\rangle\propto \delta_{n_1,n_1'}\delta_{n_2,n_2'}.
\end{align}
Next, we use the identity
\begin{align}
 \mathcal{D}^{\dagger}(\mathbf{R}) &e^{i( \bar{q} b^\dagger + q b )/\sqrt{2}} \mathcal{D}(\mathbf{R}) \nonumber \\
 &= e^{i( \bar{q} b^\dagger + q b )/\sqrt{2}} e^{i( q \bar{Z}_0 + \bar{q} Z_0 )/2} \nonumber \\
 &= e^{i\mathbf{q}\cdot\mathbf{C}} e^{i \mathbf{q}\cdot\mathbf{R}}.
 \label{eq.translation_transformation}
\end{align}
to simplify the exponential factors of the coherent state and the $\hat \Delta$ operator. Then, the expectation value is readily evaluated 
\begin{align} \label{eq.delta_matrix_element}
 \langle \mathbf{R}_{n_1,m_1}&|\hat{\Delta}_{n_1,n_2}|\mathbf{R}_{n_2,m_2}\rangle \nonumber \\
 &= \langle \mathbf{R}_{0,m_1+n_1}| e^{i(\bar{q} b^\dagger + q b)/\sqrt{2}} |\mathbf{R}_{0,m_2+n_2}\rangle \nonumber \\
 &= F_{m_1+n_1}^{m_2+n_2}(\bar{q}) e^{i \mathbf{q}\cdot\mathbf{R}}.
\end{align}
where we used Eq.~\eqref{eq.form} with the replacement $n_i\to m_i$, $a\to b$, and $\mathord{q}\to \bar{\mathord{q}}$. The inter-Landau level coherent Wigner crystal state of standard Landau levels is constructed of
\begin{align}
 |\vect{R}_{\alpha}\rangle\equiv \cos(\alpha)|\vect{R}_{0,0}\rangle+\sin(\alpha) |\vect{R}_{1,0}\rangle,
\end{align}
where we take $|R_{1,0}\rangle$ instead of the coherent state with the smallest angular momentum $|R_{1,-1}\rangle$. 
Finally, the expectation value of $\hat \Delta$ in the Wigner crystal is
\begin{align}\label{eq.WC_2} 
 \langle\text{WC}_{\alpha}&|\hat{\Delta}_{n_1,n_2}(\mathbf{q})|\text{WC}_{\alpha}\rangle \nonumber \\
 &= \begin{pmatrix}
 \cos^2\alpha F_{0,0}(\bar{q}) & \frac{1}{2}\sin 2\alpha F_{0,1}(\bar{q}) \\
 \frac{1}{2}\sin 2\alpha F_{1,0}(\bar{q}) & \sin^2\alpha F_{1,1}(\bar{q}) 
 \end{pmatrix}_{n_1,n_2} \nonumber \\
 &\quad \times \sum_{j} e^{i\mathbf{q}\cdot \mathbf{R}_j}.
\end{align}
The generalization to BLG LL is performed by replacing form factors with BLG form factors $F\to{\cal F}$.


%

\end{document}